# Distributed Object Medical Imaging Model

Ahmad Shukri Mohd Noor[1] and Md Yazid Md Saman[2.]
Department of Computer Science ,Faculty of Science and Technology
University Malaysia Terengganu
21030 Kuala Terengganu, Malaysia
*096683159 / 09-6694660(FAX)*
E-mail address: ashukri@umt.edu.my [1] / yazid@umt.edu.my [2]

**ABSTRACT**
Abstract- Digital medical informatics and images are commonly used in hospitals today,. Because of the interrelatedness of the radiology department and other departments, especially the intensive care unit and emergency department, the transmission and sharing of medical images has become a critical issue.
 Our research group has developed a Java-based Distributed Object Medical Imaging Model(DOMIM) to facilitate the rapid development and deployment of medical imaging applications in a distributed environment that can be shared and used by related departments and mobile physiciansDOMIM is a unique suite of multimedia telemedicine applications developed for the use by medical related organizations. The applications support real-time patients' data, image files, audio and video diagnosis annotation exchanges. The DOMIM enables joint collaboration between radiologists and physicians while they are at distant geographical locations. The DOMIM environment consists of heterogeneous, autonomous, and legacy resources. The Common Object Request Broker Architecture (CORBA), Java Database Connectivity (JDBC), and Java language provide the capability to combine the DOMIM resources into an integrated, interoperable, and scalable system. The underneath technology, including IDL ORB, Event Service, IIOP JDBC/ODBC, legacy system wrapping and Java implementation are explored. This paper explores a distributed collaborative CORBA/JDBC based framework that will enhance medical information management requirements and development. It encompasses a new paradigm for the delivery of health services that requires process reengineering, cultural changes, as well as organizational changes

**KEY WORDS**
Java, CORBA, DICOM , Medical Imagine ,Medical Informatics , Distributed Object Computing.

## 1. Introduction

Digital medical images are commonly used in hospitals today, even outside the radiology department. Because of the interrelatedness of the radiology department and other departments, especially the intensive care unit and emergency department, the transmission of medical images has become a critical issue. The use of World

Wide Web and network related technologies in radiology is not new. These technologies have been used in radiology teaching files to access information in multimedia integrated picture archiving and communication systems (PACS), for teleradiology purposes . Web technology has also been used to access the images stored in a Digital Imaging and Communications in Medicine (DICOM)archive in PACS environments[1][2].

## 2. Project Background

This research develops a framework of distributed medical informatics that can be used for multimedia data exchange. framework can be expand it any distributed object oriented, collaborative applications, for example, distance learning modeling and simulation

A DOMIM system based on distributed object computing sytem. The system can be viewed as a set of object services and a set of client applications. Each client application has a defined, interactive user interface. The object services provide and manage the information for the DOMIM clients. The ultimate goal is to have a complete set of services with a single fine-grained framework. The DOMIM strategy is an approach towards a single architecture where hardware and software from multiple vendors coexist in harmony. This is achieved by categorizing information into components or services (object services) as they communicate, by passing the information via interface invocations of objects. These object services are manufactured by different vendors and can run on different computers on networks. The architecture must have certain key characteristics:

(I)   *distributed:* it must support a service object model that is distributed across a regional area over LAN and WAN networks.

(2)   *platform independent:* it must support multiple computing platforms, from mainframes to servers to desktop PCs.

(3)   *heterogeneous:* it must support all different types and classes of medical equipment and software tools from many different vendors.

(4)   *location insensitive:* it must allow components in the system to replaced, repaired, upgraded and changed without compromising its ability.

Obviously, interoperability is a key technology that allows this exchange to scale. Interoperability is also the ability to leverage and reuse system content and functionality to an end user or to another system

**IJCSI**

## 3. Approach

The distributed and heterogeneous nature of todays computing systems requires a middleware infrastructure capable of supporting a three-tier computing architecture such as Common Object Request Broker Architecture (CORBA). Business logic can be built, or existing applications encapsulated, into middle-tier components that interact with end users via standard interfaces such as web browsers and standard GUI desktops, and back-end data repositories [4].

### 3.1 Distributed Object .

Common Object Request Broker Architecture (CORBA) was introduced by OMG in 1991 to go a step beyond OMA to specify technologies for interoperable distributed OO systems. With the CORBA specification, a broad and consistent model for building distributed applications is defined :
- An object-oriented based model for developing applications
- A common application programming objects in the network to be shared by client and server applications
- A syntax to define and describe the interfaces of objects used in the environment
- Support for multiple programming languages and platforms

Therefore, CORBA model formally separates the client and server portions of the application and also logically separates an application into objects that can perform certain functions. It also provides data marshaling to send and receive data with remote or local machine applications without direct knowledge of the information source or its location. In the CORBA environment, client and server applications communicate using Object Request Broker (ORB).

### 3.2 Java Based CORBA.

The Java programming language is a strongly typed, object-oriented language that borrows heavily most of its syntax from C and C++. Java is a simple, object-oriented, distributed, interpreted, robust, secure, architecture neutral, portable, high performance, multithreaded and dynamic language. This language was primarily used for developing applets-downloadable mini-applications that could be embedded inside Web pages and performed in browsers. However, since 1995, Java has emerged as a first-class programming language that is being used for everything from embedded devices to enterprise servers. Nowadays the Java language can be seen in use in a wider range of applications. When an application is written and compiled in one place it can run on any machine under any operating system. Sometimes the "Write Once, Run Anywhere" slogan is called the synonym of Java. Anyway, platform independence is the ability of a program to move from one computer system to another. Java is platform independent at both the source and the binary level. The secret of the Java has been hidden into Java Virtual machine (JVM). Instead of creating machine dependent code, the Java compiler creates a bytecode format, which can be run on any Virtual Machine (VM). Somehow, Java makes programming easier because it is object-oriented and has automatic garbage collection.

Java offers tremendous flexibility for distributed application development. To do this, Java needs to be augmented with a distributed object infrastructure, which is where OMG's CORBA comes into the picture. Using CORBA requires more than just a knowledge of the CORBA architecture. CORBA should be part of a well designed system architecture.

CORBA technology as part of the Java 2 platform consists of an Object Request Broker (ORB) written in Java. Java IDL adds CORBA capability to the Java platform, providing standards-based interoperability and connectivity. Java IDL enables distributed Web-enabled Java applications to transparently invoke operations on remote network services using the industry standard OMG IDL (Interface Definition Language) and IIOP (Internet Inter-ORB Protocol) defined by the Object Management Group. CORBA is an distibuted object-oriented middleware protocols, used for the DOMIM development. By using CORBA it gives us several benefits in the system distributed computing environment. For example, we are able to interface legacy database by developing CORBA wrapper that allows us to access the data structures in the database without disturbing the existing database. Interoperability and scalability are other benefits of using CORBA. The CORBA IDL for streaming medical image in our model as follow :-

```
struct Info {
string name; module Student_App {

struct Info {
string name;
string matric;
string address;
string city;
string state;
string zip;
string country;
string email;
string phone;
string program;
};

typedef sequence<octet> Data;
```

![IJCSI]

```
   interface project {
      string execute(in short operation, in Info
info_student);
      Info execute2(in short operation, in Info
info_student);
      Data downloadFile(in string fileName);
   };

};
   string matric;
   string address;
   string city;
   string state;
   string zip;
   string country;
   string email;
   string phone;
   string program;
   };

   typedef sequence<octet> Data;

   interface project {
      string execute(in short operation, in Info
info_student);
      Info execute2(in short operation, in Info
info_student);
      Data downloadFile(in string fileName);
   };

};
module Student_App {

   struct Info {
   string name;
   string matric;
   string address;
   string city;
   string state;
   string zip;
   string country;
   string email;
   string phone;
   string program;
   };

   typedef sequence<octet> Data;

   interface project {
      string execute(in short operation, in Info
info_student);
      Info execute2(in short operation, in Info
info_student);
      Data downloadFile(in string fileName);
```

};

```
   typedef sequence<octet> Data;
   interface project {
   string execute(in short operation, in Info info_patient);
   Info execute2(in short operation, in Info info_patient);
   Data downloadFile(in string fileName);
```

### 3.3 Digital Imaging and Communications in Medicine

The Digital Imaging and Communications in Medicine (DICOM) standard was created by the National Electrical Manufacturers Association (NEMA) to aid the distribution and viewing of medical images, such as CT scans and ultrasound. New technologies such as Java should always be used as complements of the de facto standard in medical imagine, DICOM. DICOM allows the interchange of images from different modalities, archives, and workstations from different vendors. java technology can be used to build a storage system and to make this service accessible for different clients. However, this storage service should also incorporate DICOM services to store and access examination data from DICOM workstations and DICOM modalities. Since Java version 1.4, the Java standard includes a specification for working with images stored in files and accessed across the network. This specification is called Java Image I/O. It provides a pluggable framework for easily adding support for alternate image formats using third-party plug-ins. The DICOM Image I/O Plug-in connects the DICOM® standard to the Java™ standard. DICOM is the universal standard for sharing medical imaging resources between heterogeneous and multi-vendor equipments (acquisition device, workstation, storage server, patient management system, etc.).

### 3.4 Distributed Medical API Impelememtaion

A toolkit, which is referred to as *NeatMed*, is intended to reduce development time by eliminating the need for the application developer to deal directly with medical image data The medical imaging application developers interface(API), NeatMed interface (API), was developed using the Java programming language (Sun Microsystems, Mountain View, Calif). An extension API is a set of classes that can be instantiated by a programmer to create a particular type of application, thus facilitating software reuse. NeatMed is an example of an extension API that can be used for the development of applications that deal with off-line medical image data..NeatMed currently provides support for the Digital Imaging and Communications in Medicine and Analyze medical image file formats. The NeatMed API is a group of core and support classes that can be used to interpret, represent, and manipulate images and related data that are stored in DICOM-compliant files. The toolkit has been created specifically for interpreting medical image data; it thus acts as a platform for development of medical imaging applications.. NeatMed was implemented by using Java[1]. NeatMed currently provides support for the DICOM in Medicine and Analyze medical image file formats.The Neat-Med API is distributed in accordance with the terms and conditions laid out in the GNU Lesser General Public License[2]. This license was selected to ensure that the NeatMed. NeatMed was developed using the Java programming languages. It was initially intended for the development of software for use in the consumer electronics industry (eg, set-top boxes). The core Java libraries maintained by Sun Microsystems are used as the foundation for the development of any Java application. These libraries can be used in conjunction with an extension API in order to develop specialized applications. An extension API is a set of classes that can be instantiated by a programmer to create a particular type of application, thus facilitating software reuse. NeatMed is an example of an extension API that can be used for the development of applications that deal with off-line medical image data. A large number of Java APIs exist; these deal with a broad range of applications ranging from communicating with the serial and parallel ports to advanced image processing. The set of classes representing the API is deployed in some type of library. Java provides a packaging tool that can be used to package a set of class files and associated resources into a Java archive or *JAR* file. In order to be useful, an API must be well documented. Java provides a documentation tool called *Javadoc* that allows an API developer to document software as it is being written. The resulting documentation provides detailed information about each class, method, and variable that is defined in the associated API. The structure of Javadoc documentation is more or less the same for every API. This makes it very easy for programmers to familiarize themselves with a new API once they are comfortable with the basic Javadoc documentation structure. The API documentation is generated in HTML and can be viewed using any standard Web browser. Java has a wide range of benefits associated with it; however, there are also some limitations. One example is performance: Java is a multiplatform programming language; the byte code (ie, binary form) that represents a Java program is interpreted and not executed directly. This reduces the performance of a Java program compared to a natively executed program. Overall, however, the benefits associated with Java (listed previously) far outweigh the drawbacks, hence its selection for the development of NeatMed.. In

Fig 2, Client Programming structure show flexibility, and ease of use of the NeatMed API in java Environment

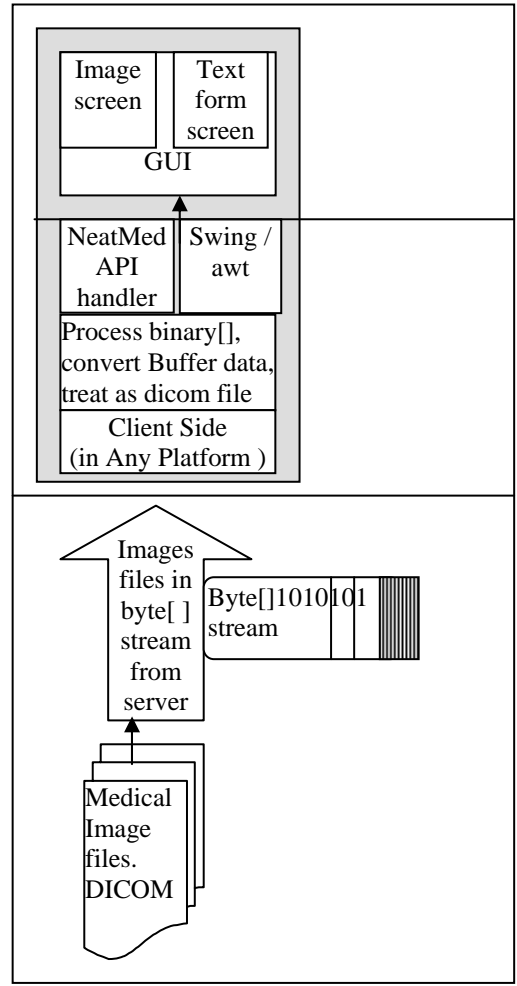

**Figure 2: Client Programming structure.**

The central class in the API is the DICOMImage class. A DICOMImage object can be instantiated by specifying a reference to a suitable data source in the constructor. The constructor will accept data from a number of sources (eg, local file, data stream, and remote uniform resource locator (URL)). Once constructed, a DICOMImage object provides direct access to all of the data elements stored within the specified DICOM source. Other classes in the API are used to represent individual components within a DICOM RadioGraphics.

46

Here is the example :

*DICOMImagePlus image = new DICOMImagePlus(ss);*
*JLabel label = new JLabel(new ImageIcon(image.getAsBufferedImage(0)));*
*JFrame frame = new JFrame("Patient NO : " +Patient_info2.id +" NAME : " + Patient_info2.name + " Image File :"+ Patient_info2.file_id);*
*frame.getContentPane().setLayout(new BorderLayout());*
*frame.getContentPane().add(label);*
*frame.setVisible(true);*
*frame.pack();}*

The choice of Java for implementing the NeatMed API was also influenced by a number of its key features:

- Ease of use: Java is a modern programming language that was designed with simplicity in mind. Many of the complexities that are associated with other programming languages have been omitted, whereas much of the power and flexibility has been retained. This makes Java very easy to learn and use, particularly in the case of novice programmers.
- Level of support: Although Java is a relatively new programming language, there is a significant amount of support material available. Numerous texts have been written dealing with all aspects of the language. In addition, tutorials, sample source code, API documentation, and freely available integrated development environments (IDEs) can all be accessed via the Internet.
- Portability: Java is a multiplatform programming language. This means that a Java

**5. Distributed Medical informatics Architecture.**

The distributed medical informatics architecture design includes client applications at the 1st tier that access remote medical imaging data query server and database at 3rd tier via Java ORB as a middle agent at 2nd tier,

The client's implementations comprised of a Java application . All the user interfaces were created using Java APIs while NeatMed API for presenting and displaying the medical image files in DICOM format.

The server is comprised of the some logical algorithms that responsible for executing an input query statement from the client and returning the query results back to the client. The connection between the servers and the IBM DB2 Database is accomplished via its native JDBC driver. This server is placed on the local area network (LAN) with Java ORB acting as the middleware. The ORB using octet-streaming services is utilized to transfer multimedia data such as medical images, It also can be used for audio and video in a 3-tiers heterogeneous environment.

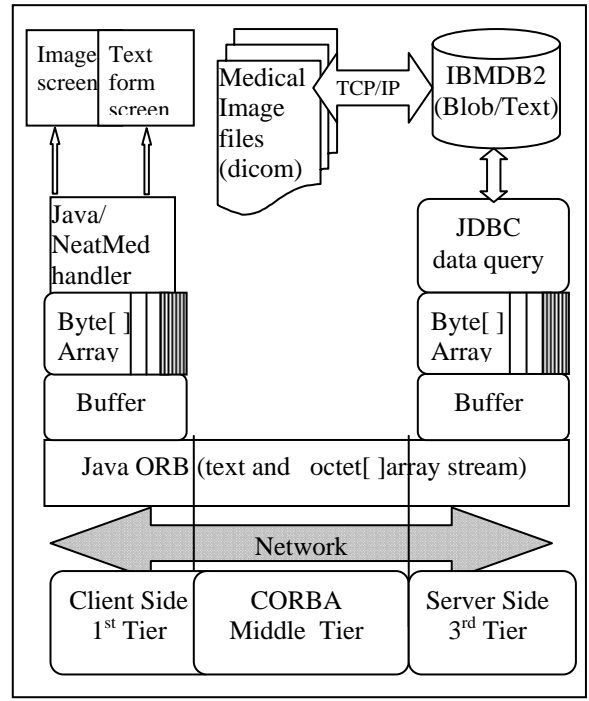

**Figure3: Distributed Object Medical Imaging Model (DOMIM) Architecture**

The back-end tier of the architecture involves the storage and retrieval of multimedia data on the database server.
In this paper, the Object Relational database management system (ORDBMS) is used for the development of medical imaging and multimedia database as it allows queries to be performed on complex data, e.g. images,video, audio, etc. The following components are utilised in the development of the server application:
- IBM DB2 v 8.1 PE database.
- Server application (ProjectServer.class) for receiving object and sending back the object from/to client application.
- Server application logic (ProjectImpl.class) for executing as input query

IBM DB2 has been chosen as the implementation database. Due to IBM DB2 ability to support the binary large object blocks (BLOBs) data. All the image data are stored in their native binary format in a particular column of the database table.

The following features are provided by the client's implementations(ProjectClient.class) via a Graphical User Interface (GUI):
- Binding to the servers' implementations.
- Invoking the servers' implementation with the appropriate commands.
- Displaying and presenting the query results(text and images) to the user.

The design depicts that ORB can be an integral part of deploying Java applets/applications, including those that access database. The diagram also shows that the client application does not directly connect to the databases. Instead, Java ORB facilitates database connectivity by allowing client-side objects to communicate with server-side objects that assume responsibility for performing database access. That is, server-side objects is written to handle database access on behalf of a client object(s) that instantiated it. The client main function is to display multiple media data to the user in the specified format. Such architecture can provide adequate database support for medical applications demanding interactive medical imaging presentations .the modules involved are:-

- GUI Object Class

The Graphical User Interface (GUI) class represents the tools used for human-to-computer interaction . The GUI may include menus, buttons, graphics, textual and visual information, image annotation, full-duplex audio, and video information. The GUI class is responsible for collecting relevant information to be passed onto the user. The GUI class also distributes user input to objects whose states depend on the user. The objects in the class are linked with other objects in the system , e.g., management and control objects.

- Audio Object Class

The audio object class handles real-time and playback of audio sequences at the user workstation. Digital audio is encoded to sequences of 8-bit or 16-bit samples using standard telephone Coder/Decoder (CODEC) conversion. The audio sequences are two-way conversations between physicians or audio notes to be stored with a patient record.

- Video Object Class

The video object class handles real-time and playback video sequences. The video sequences are digital format using MPEG compression frames. Video sequences can be generated from digital camera or pre-recorded video on CD disk formats. Video sequences can be stored as video object classes in the multimedia database archive system.

- Buffer Object Class

The buffer object class controls jitter by queuing data until it is synchronized and ready for use. Buffering takes place on all data traversing the network stack to abstract the application from network timing idiosyncrasies and dependencies.

- File Stream Object Class

The file stream objects include files, e.g., images, which must be transferred between nodes over networks. These nodes can be workstations or the Database Archive System. The objects represent a flow of packets, and the objects inherent the format of their types, i.e., audio, video, annotation commands, etc.

- Communication Object Class

The communication object class represents a group of connectivity mechanism. For example, objects conduct communication via middleware APIs or TCP/IP sockets. These objects provide the communication paradigm to the data objects described above, i.e., annotation commands, file streams, etc. and link the multimedia information exchanging over the distributed computing environment.

## 6. Implemantation

In this application, client patient's detail screen and medical image in DICOM format screen are displyed in separated windows.

i. The client request patient data by providing patient case retrieving procedures such as patient id and notifies the server(services Perovider) via corba event service .

ii. The Server retrieve a patient demographic data and image(s) from client and send the DBA via JDBC Server using JDBC connectivity to the DBMS, e.g., IBM DB2 database. As in figure 5

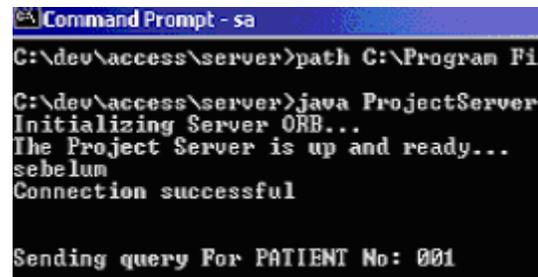

**Fig 6:Server response screen shot**

iii. The Server passes the patient demographic data and image file identification (where those files *resid ein remote storage) to the client via ORB.*

iv. The remote Client fetches the demographic data and display then details on client GUI screen as depict in figure 6 . Then it point the patient image(s) based on the given file identification. then, the system pop-up the medical image as illustrated figure 7 below

48

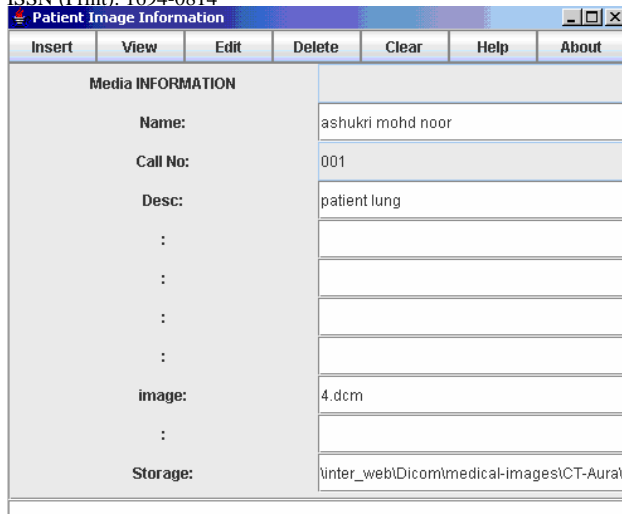

Fig 7. Screenshot of GUI patient's data

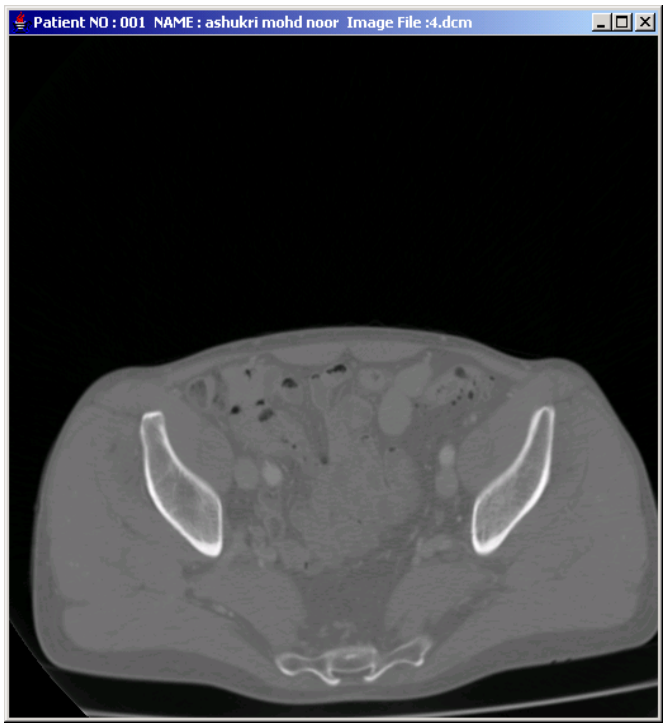

Fig 8. Screenshot of patient's Medical Image

## 7. Conclusion

The three-tier distributed medical imagine application allow the application interoperability and independence of platform, operating system, programming language and even of network and protocolThe application architecture or framework is an important and common stage in the development of any medical imaging application in distributed environment.

NeatMed removes the need to deal directly with encoded medical image data, thus increasing productivity and allowing the developer to concentrate on other aspects of application development. NeatMed is written in Java, a multiplatform programming language with a large amount of freely available support material that is straightforward to learn and use.These and other features of Java make NeatMed accessible to a large group of potential users. Most important, NeatMed is a freely available research tool whose ongoing development is driven by the needs and requirements of its users.